# Structural analysis of SARS-CoV-2 and predictions of the human interactome

Andrea Vandelli[1,2], Michele Monti[1,3], Edoardo Milanetti[4,5],
Jakob Rupert [6], Elsa Zacco[3], Elias Bechara[1,3],
Riccardo Delli Ponti [7,*] and Gian Gaetano Tartaglia [1,3,6,8,*]

[1] Centre for Genomic Regulation (CRG), The Barcelona Institute for Science and Technology, Dr. Aiguader 88, 08003 Barcelona, Spain and Universitat Pompeu Fabra (UPF), 08003 Barcelona, Spain

[2] Systems Biology of Infection Lab, Department of Biochemistry and Molecular Biology, Biosciences Faculty, Universitat Autònoma de Barcelona, 08193 Cerdanyola del Vallès, Spain

[3] Department of Neuroscience and Brain Technologies, Istituto Italiano di Tecnologia, Via Morego 30, 16163, Genoa, Italy.

[4] Department of Physics, Sapienza University, Piazzale Aldo Moro 5, 00185, Rome, Italy

[5] Center for Life Nanoscience, Istituto Italiano di Tecnologia, Viale Regina Elena 291, 00161, Rome, Italy

[6] Department of Biology 'Charles Darwin', Sapienza University of Rome, P.le A. Moro 5, Rome 00185, Italy

[7] School of Biological Sciences, Nanyang Technological University, 60 Nanyang Drive, Singapore, 637551, Singapore

[8] Institucio Catalana de Recerca i Estudis Avançats (ICREA), 23 Passeig Lluis Companys, 08010 Barcelona, Spain

*to whom correspondence should be addressed to: riccardo.ponti@ntu.edu.sg (RDP) and giangaetano.tartaglia@uniroma1.it or gian.tartaglia@iit.it (GGT)

## ABSTRACT

Specific elements of viral genomes regulate interactions within host cells. Here, we calculated the secondary structure content of >2500 coronaviruses and computed >100000 human protein interactions with severe acute respiratory syndrome coronavirus 2 (SARS-CoV-2). We found that the 3' and 5' are the most structured elements in the viral genome and the 5' has the strongest propensity to associate with human proteins. The domain encompassing nucleotides 23000 – 24000 is highly conserved both at the sequence and structural level, while the region upstream varies significantly. These two sequences code for a domain of the viral protein Spike S that interacts with the human receptor angiotensin-converting enzyme 2 (ACE2) and has the potential to bind sialic acids. Our predictions indicate that the first 1000 nucleotides in the 5' can interact with proteins involved in viral RNA processing such as double-stranded RNA specific editases and ATP-dependent RNA-helicases, in addition to other high-confidence candidate partners. These interactions, previously reported to be also implicated in HIV, reveal important information on host-virus interactions. The list of transcriptional and post-transcriptional elements recruited by SARS-CoV-2 genome provides clues on the biological pathways associated with gene expression changes in human cells.

**INTRODUCTION**

A disease named Covid-19 by the World Health Organization and caused by the severe acute respiratory syndrome coronavirus 2 (SARS-CoV-2) has been recognized as responsible for the pneumonia outbreak that started in December, 2019 in Wuhan City, Hubei, China [1] and spread in February to Milan, Lombardy, Italy [2] becoming pandemic. As of April 2020, the virus infected >2’000’000 people in >200 countries.

SARS-CoV-2 is a positive-sense single-stranded RNA virus that shares similarities with other beta-coronavirus such as severe acute respiratory syndrome coronavirus (SARS-CoV) and Middle East respiratory syndrome coronavirus (MERS-CoV) [3]. Bats have been identified as the primary host for SARS-CoV and SARS-CoV-2 [4,5] but the intermediate host linking SARS-CoV-2 to humans is still unknown, although a recent report indicates that pangolins could be involved [6].

Coronaviruses use species-specific proteins to mediate the entry in the host cell and the spike S protein activates the infection in human respiratory epithelial cells in SARS-CoV, MERS-CoV and SARS-CoV-2 [7]. Spike S is assembled as a trimer and contains around 1,300 amino acids within each unit [8,9]. The receptor binding domain (RBD) of Spike S, which contains around 300 amino acids, mediates the binding with angiotensin-converting enzyme, (ACE2) attacking respiratory cells. Another region upstream of the RBD, present in MERS-CoV but not in SARS-CoV, is involved in the adhesion to sialic acid and could play a key role in regulating viral infection [7,10].

At present, very few molecular details are available on SARS-CoV-2 and its interactions with the human host, which are mediated by specific RNA elements [11]. To study the RNA structural content, we used *CROSS* [12] that was previously developed to investigate large transcripts such as the human immunodeficiency virus HIV-1 [13]. *CROSS* predicts the structural profile of RNA molecules (single-and double-stranded state) at single-nucleotide resolution using sequence information only. Here, we performed sequence and structural alignments among 62 SARS-CoV-2 strains and identified the conservation of specific elements in the spike S region, which provides clues on the evolution of domains involved in the binding to ACE2 and sialic acid.

As highly structured regions of RNA molecules have strong propensity to form stable contacts with proteins [14] and promote assembly of specific complexes [15,16], SARS-CoV-2 domains enriched in double-stranded content are expected to establish interactions within host cells that are important to replicate the virus [17]. To investigate the interactions of SARS-CoV-2 RNA with human proteins, we employed *cat*RAPID [18,19]. *cat*RAPID [20] estimates the binding potential of a specific protein for an RNA molecule through van der Waals, hydrogen bonding and secondary structure propensities allowing identification of interaction partners with high confidence [21]. The unbiased analysis of more than 100000 protein interactions with SARS-CoV-2 RNA reveals that the 5' of SARS-CoV-2 has strong propensity to bind to human proteins involved in viral infection and especially reported to be associated with HIV infection. A comparison between SARS-CoV and HIV reveals indeed similarities [22], but the relationship between SARS-CoV-2 and HIV is still unexplored. Interestingly, HIV and SARS-CoV-2, but not SARS-CoV nor MERS-CoV, have a furin-cleavage site occurring in the spike S protein, which could explain the spread velocity of SARS-CoV-2 compared to SARS-CoV and MERS-CoV [23,24]. Yet, many processes related to SARS-CoV-2 replication are unknown and our study aims to suggest relevant protein interactions for further investigation.

We hope that our large-scale calculations of structural properties and binding partners of SARS-CoV-2 will be useful to identify the mechanisms of virus replication within the human host.

## RESULTS

### SARS-CoV-2 contains highly structured elements

Structured elements within RNA molecules attract proteins [14] and reveal regions important for interactions with the host [25]. Indeed, each gene expressed from SARS-CoV-2 is preceded by conserved transcription-regulating sequences that act as signal for the transcription complex during the synthesis of the RNA minus strand to promote a strand transfer to the leader region to resume the synthesis. This process is named discontinuous extension of the minus strand and is a variant of similarity-assisted template switching that operates during viral RNA recombination [17].

To analyze SARS-CoV-2 structure (reference Wuhan strain MN908947.3), we employed *CROSS* [12] to predict the double- and single-stranded content of RNA genomes such as HIV-1 [13]. We found the highest density of double-stranded regions in the 5' (nucleotides 1-253), membrane M protein (nucleotides 26523-27191), spike S protein (nucleotides 23000-24000), and nucleocapsid N protein (nucleotides 2874-29533; **Fig. 1**) [26]. The lowest density of double-stranded regions were observed at nucleotides 6000-6250 and 20000-21500 and correspond to the regions between the non-structural proteins nsp14 and nsp15 and the upstream region of the spike surface protein S (**Fig. 1**) [26]. In addition to the maximum corresponding to nucleotides 23000-24000, the structural content of spike S protein shows minima at around nucleotides 20500 and 24500 (**Fig. 1**).
We used the *Vienna* method [27] to further investigate the RNA secondary structure of specific regions identified with *CROSS* [13]. Employing a 100 nucleotide window centered around *CROSS* maxima and minima, we found good match between *CROSS* scores and Vienna free energies (**Fig. 1).** Strong agreement is also observed between *CROSS* and *Vienna* positional entropy, indicating that regions with the highest structural content have also the lowest structural diversity.

Our analysis suggests the presence of structural elements in SARS-CoV-2 that have evolved to interact with specific human proteins [11]. Our observation is based on the assumption that structured regions have an intrinsic propensity to recruit proteins [14], which is supported by the fact that structured transcripts act as scaffolds for protein assembly [15,16].

## Structural comparisons reveal that the spike S region of SARS-CoV-2 is conserved among coronaviruses

We employed *CROSS*align [13] to study the structural conservation of SARS-CoV-2 in different strains (**Materials and Methods**).

In our analysis, we compared the Wuhan strain MN908947.3 with around 2800 other coronaviruses (data from NCBI) with human (**Fig. 2**) or other hosts (**Supp. Fig. 1**). When comparing SARS-CoV-2 with human coronaviruses (1387 strains, including SARS-CoV and MERS-CoV), we found that the most conserved region falls inside the spike S genomic locus (**Fig. 2**). More precisely, the conserved region is between nucleotides 23000 - 24000 and exhibits an intricate and stable secondary structure (RNA*fold* minimum free energy= -269 kcal/mol )[27]. High conservation of a structured region suggests a functional activity that might be relevant for host infection.

The 3' and 5' of SARS-CoV-2 are highly variable in our set. However, the 3' and 5' are more structured in SARS-CoV-2 than other coronaviruses (average structural content for SARS-CoV-2 = 0.56, indicating that 56% of the CROSS signal is >0, in the 5' and 0.49 in the 3'; by contrast, other coronaviruses have 0.49 in the 5' and 0.42 in the 3').

**Sequence and structural comparisons among SARS-CoV-2 strains indicate conservation of the ACE2 binding site and high variability in a region potentially interacting with sialic acids.**

To better investigate the sequence conservation of SARS-CoV-2, we compared 62 strains isolated from different countries during the pandemic (including China, USA, Japan, Taiwan, India, Brazil, Sweden, and Australia; data from NCBI and in VIPR www.viprbrc.org; **Materials and Methods**). Our analysis aims to determine the relationship between structural content and sequence conservation.

Using *Clustal W* for multiple sequence alignments [28], we observed general conservation of the coding regions (**Fig. 3A**). The 5' and 3' show high variability due to experimental procedures of the sequencing and are discarded in this analysis [29] . One highly conserved region is between nucleotides 23000 - 24000 in the spike S genomic locus, while sequences up- and downstream are variable (red bars in **Fig. 3A**). We then used CROSS*align* [13] to compare the structural content (**Materials and Methods**). High variability of structure is observed for both the 5' and 3' and for nucleotides between 21000 - 22000 as well as 24000 - 25000, associated with the S region (red bars in **Fig. 3A**). The rest of the regions are significantly conserved at a structural level (p-value < 0.0001; Fisher's test).

We then compared protein sequences coded by the spike S genomic locus (NCBI reference QHD43416) and found that both sequence (**Fig. 3A**) and structure (**Fig. 2**) of nucleotides 23000 - 24000 are highly conserved. The region corresponds to amino acids 330-500 that contact the host receptor angiotensin-converting enzyme 2 (ACE2) [30] promoting infection and provoking lung injury [24,31]. By contrast, the region upstream of the binding site receptor ACE2 and located in correspondence to the minimum of the structural profile at around nucleotides 22500-23000 (**Fig. 1**) is highly variable [32], as indicated by *T-coffee* multiple sequence alignments [32] (**Fig. 3A**). This part of the spike S region corresponds to amino acids 243-302 that in MERS-CoV binds to sialic acids regulating infection through cell-cell membrane fusion (**Fig. 3B;** see related manuscript by E. Milanetti *et al.*) [10,33,34].

Our analysis suggests that the structural region between nucleotides 23000 and 24000 of Spike S region is conserved among coronaviruses (**Fig. 2**) and that the binding site for ACE2 has poor variation in human SARS-CoV-2 strains (**Fig. 3B**). By contrast, the region upstream, which has propensity to bind sialic acids [10,33,34], showed poor structural content and high variability (**Fig. 3B**).

**Analysis of human interactions with SARS-CoV-2 identifies proteins involved in viral replication**

In order to obtain insights on how the virus replicates in human cells, we predicted SARS-CoV-2 interactions with the whole RNA-binding human proteome. Following a protocol to study structural conservation in viruses [13], we first divided the Wuhan sequence in 30 fragments of 1000 nucleotides each moving from the 5' to 3' and then calculated the protein-RNA interactions of each fragment with *cat*RAPID *omics* (3340 canonical and putative RNA-binding proteins, or RBPs, for a total 102000 interactions) [18]. Proteins such as Polypyrimidine tract-binding protein 1 PTBP1 (Uniprot P26599) showed the highest interaction propensity (or Z-score; **Materials and Methods**) at the 5' while others such as Heterogeneous nuclear ribonucleoprotein Q HNRNPQ (O60506) showed the highest interaction propensity at the 3', in agreement with previous studies on coronaviruses (**Fig. 4A**) [35].

For each fragment, we predicted the most significant interactions by filtering according to the Z score. We used three different thresholds in ascending order of stringency: $Z \geq 1.50$, 1.75 and 2 respectively and we removed from the list the proteins that were predicted to interact promiscuously with more than one fragment. Fragment 1 corresponds to the 5' and is the most contacted by RBPs (around 120 with $Z \geq 2$ high-confidence interactions; **Fig. 4B**), which is in agreement with the observation that highly structured regions attract a large number of proteins [14]. Indeed, the 5' contains multiple stem loop structures that control RNA replication and transcription [36,37]. By contrast, the 3' and fragment 23 (Spike S), which are still structured but to a lesser extent, attract fewer proteins (10 and 5, respectively), while fragment 20 (between orf1ab and Spike S) that is predicted to be unstructured, does not have binding partners.

The interactome of each fragment was then analysed using *clever*GO, a tool for Gene Ontology (GO) enrichment analysis [38]. Proteins interacting with fragments 1, 2 and 29 were associated with annotations related to viral processes (**Fig. 4C; Supp. Table 1**). Considering the three thresholds applied (**Materials and Methods**), we found 22 viral proteins for fragment 1, 2 proteins for fragment 2 and 11 proteins for fragment 29 (**Fig. 4D**).

Among the high-confidence interactors of fragment 1, we discovered RBPs involved in positive regulation of viral processes and viral genome replication, such as double-stranded RNA-specific editase 1 ADARB1 (Uniprot P78563 [39]), 2-5-oligoadenylate synthase 2 OAS2 (P29728) and 2-5A-dependent ribonuclease RNASEL (Q05823). Interestingly, 2-5-oligoadenylate synthase 2 OAS2 has been reported to be upregulated in human alveolar adenocarcinoma (A549) cells infected with SARS-CoV-2 (log fold change of 4.2; p-value of $10^{-9}$ and q-value of $10^{-6}$) [40]. While double-stranded RNA-specific adenosine deaminase ADAR (P55265) is absent in our library due to its length that does not meet *cat*RAPID *omics* requirements [18], the *omiXcore* extension of the algorithm specifically developed for large molecules [41] attributes the same binding propensity to both ADARB1 and ADAR, thus indicating that the interactions might occur (**Materials and Methods**). Moreover, experimental works indicate that the family of ADAR deaminases is active in bronchoalveolar lavage fluids derived from SARS-CoV-2 patients [42] and is upregulated in A549 cells infected with SARS-CoV-2 (log fold change of 0.58; p-value of $10^{-8}$ and q-value of $10^{-5}$) [40].

We also identified proteins related to the establishment of integrated proviral latency, including X-ray repair cross-complementing protein 5 XRCC5 (P13010) and X-ray repair cross-complementing protein 6 XRCC6 (P12956; **Fig. 4B** and **4E**). In accordance with our calculations, comparison of A549 cells responses to SARS-CoV-2 and respiratory syncytial virus, indicates upregulation of XRRC6 in SARS-CoV-2 (log fold-change of 0.92; p-value of 0.006 and q-value of 0.23) [40]. Nucleolin NCL (P19338), a protein known to be involved in coronavirus processing, was also predicted to bind tightly to the 5' (**Supp. Table 1**) [43].

Importantly, we found proteins related to defence response to viruses, such as ATP-dependent RNA helicase DDX1 (Q92499), that are involved in negative regulation of viral genome replication. Some DNA-binding proteins such as Cyclin-T1 CCNT1 (O60563), Zinc finger protein 175 ZNF175 (Q9Y473) and Prospero homeobox protein 1 PROX1 (Q92786) were included because they could have potential RNA-binding ability (**Fig. 4E**) [44]. As for fragment 2, we found two canonical RBPs:

E3 ubiquitin-protein ligase TRIM32 (Q13049) and E3 ubiquitin-protein ligase TRIM21 (P19474), which are listed as negative regulators of viral release from host cell, negative regulators of viral transcription and positive regulators of viral entry into host cells. Among these genes, DDX1 (log fold change of 0.36; p-value of 0.007 and q-value of 0.23) and TRIM21 (log fold change of 0.44; p-value of 0.003 and q-value of 0.18) are the most significantly upregulated in A549 cells infected with SARS-CoV-2 [40]. Finally, 10 of the 11 viral proteins detected for fragment 29 are members of the Gag polyprotein family, that perform different tasks during HIV assembly, budding, maturation. More than simple scaffold elements, Gag proteins are versatile elements that bind to viral and host proteins as they traffic to the cell membrane (**Supp. Table 1**) [45].

Analysis of functional annotations carried out with *GeneMania* [46] revealed that proteins interacting with the 5' of SARS-CoV-2 RNA are associated with regulatory pathways involving NOTCH2, MYC and MAX that have been previously connected to viral infection processes (**Fig. 4E**) [47,48]. Interestingly, some proteins, including DDX1, CCNT1 and ZNF175 for fragment 1 and TRIM32 for fragment 2, have been shown to be necessary for HIV functions and replication inside the cell. The roles of these proteins in retroviral replication are expected to be different in SARS-CoV-2. As SARS-CoV-2 represses host gene expression through a number of unknown mechanisms [4849], sequestration of elements such as DDX1 and CCNT1 could be exploited to alter biological pathways in the host cell. Supporting this hypothesis, DXX1 and CCNT1 have been shown to condense in large ribonucleoprotein assemblies such as stress granules [50–52] that are the primary target of RNA viruses [53]. Regarding the function of these proteins in disease, DDX1 is required for HIV-1 Rev as well as for avian coronavirus IBV replication and it binds to the RRE sequence of HIV-1 RNAs [1–3], while CCNT1 binds to 7SK snRNA and regulates transactivation domain of the viral nuclear transcriptional activator, Tat [56,57].

**Analysis of SARS-CoV-2 proteins reveals common interactions**

Recently, Gordon *et al.* reported a list of human proteins binding to Open Reading Frames (ORFs) translated from SARS-CoV-2 [58]. Identified through affinity purification followed by mass spectrometry quantification, 332 proteins from HEK-293T cells interact with viral ORF peptides. By selecting 274 proteins binding at the 5' with Z score ≥1.5 (**Supp. Table 1**), of which 140 are exclusively interacting with fragment 1 (**Fig. 4B**), we found that 8 are also reported in the list by Gordon *et al.* [58], which indicates significant enrichment (representation factor of 2.5; p-value of 0.02; hypergeometric test with human proteome in background). The fact that our list of protein-RNA binding partners contains elements identified also in the protein-protein network analysis is not surprising, as ribonucleoprotein complexes evolve together [14] and their components sustain each other through different types of interactions [16].

We note that out of 332 interactions, 60 are RBPs (as reported in Uniprot [39]), which represents a considerable fraction (i.e., 20%), considering that there are around 1500 RBPs in the human proteome (i.e., 6%) and fully justified by the fact that they involve association with viral RNA. Comparing the RBPs present in Gordon *et al.* [58] and those present in our list (79 as in Uniprot), we found an overlap of 6 proteins (representation factor = 26.5; p-value < $10^{-8}$; hypergeometric test), including: Janus kinase and microtubule-interacting protein 1 JAKMIP1 (Q96N16), A-kinase anchor protein 8 AKAP8 (O43823) and A-kinase anchor protein 8-like AKAP8L (Q9ULX6), which in case of HIV-1 infection is involved as a DEAD/H-box RNA helicase binding [59], signal recognition particle subunit SRP72 (O76094), binding to the 7S RNA in presence of SRP68, La-related protein 7, LARP7 (Q4G0J3) and La-related protein 4B LARP4B (Q92615), which are part of a system for transcriptional regulation acting by means of the 7SK RNP system [60] (**Fig. 4F; Supp. Table 2**). Interestingly, SRP72, LARP7 and LARP4B proteins assemble in stress granules [61–63] that are the targeted by RNA viruses [53]. We speculate that sequestration of these elements is orchestrated by a viral program aiming to recruit host genes [49]. LARP7 is also upregulated in A549 cells infected with SARS-CoV-2 (log fold change of 0.48; p-value of 0.006 and q-value of 0.23) [40].

Moreover, by directly analysing the RNA interaction potential of all the 332 proteins by Gordon *et al.* [58], *cat*RAPID identified 38 putative binders at the 5’ (Z score ≥ 1.5; 27 occurring exclusively in the 5’ and not in other regions of the RNA) [18], including Serine/threonine-protein kinase TBK1 (Q9UHD2), among which 10 RBPs (as reported in Uniprot) such as: Splicing elements U3 small nucleolar ribonucleoprotein protein MPP10 (O00566) and Pre-mRNA-splicing factor SLU7 (O95391), snRNA methylphosphate capping enzyme MEPCE involved in negative regulation of transcription by RNA polymerase II 7SK (Q7L2J0) [64], Nucleolar protein 10 NOL10 (Q9BSC4) and protein kinase A Radixin RDX (P35241; in addition to those mentioned above; **Supp. Table 2**).

## HIV-related RBPs are significantly enriched in the 5’ interactions

In the list of 274 proteins binding to the 5’ (fragment 1) with Z score ≥1.5, we found 10 hits associated with HIV (**Supp. Table 3**), which represents a significant enrichment (p-value=0.0004; Fisher’s exact test), considering that the total number of HIV-related proteins is 35 in the whole

*cat*RAPID library (3340 elements). The complete list of proteins includes ATP-dependent RNA helicase DDX1 (Q92499 also involved in Coronavirus replication [54,55]), ATP-dependent RNA helicase DDX3X (O00571 also involved in Dengue and Zika Viruses), Tyrosine-protein kinase HCK (P08631, nucleotide binding), Arf-GAP domain and FG repeat-containing protein 1 (P52594), Double-stranded RNA-specific editase 1 ADARB1 (P78563), Insulin-like growth factor 2 mRNA-binding protein 1 IGF2BP1 (Q9NZI8), A-kinase anchor protein 8-like AKAP8L (Q9ULX6; its partner AKAP8 is also found in Gordon *et al.* [58]) Cyclin-T1 CCNT1 (O60563; DNA-binding) and Forkhead box protein K2 FOXK2 (Q01167; DNA-binding; **Fig. 4B,E Supp. Table 3**).

Smaller enrichments were found for proteins related to Hepatitis B virus (HBV; p-value=0.01; 3 hits out of 7 in the whole *cat*RAPID library; Fisher's exact test), Nuclear receptor subfamily 5 group A member 2 NR5A2 (DNA-binding; O00482), Interferon-induced, double-stranded RNA-activated protein kinase EIF2AK2 (P19525), and SRSF protein kinase 1 SRPK1 (Q96SB4) as well as Influenza (p-value=0.03; 2 hits out of 4; Fisher's exact test), Synaptic functional regulator FMR1 (Q06787) and RNA polymerase-associated protein RTF1 homologue (Q92541; **Supp. Table 3**). By contrast, no significant enrichments were found for other viruses such as for instance Ebola.

Very importantly, specific chemical compounds have been developed to interact with HIV- and HVB-related proteins. The list of HIV-related targets reported in ChEMBL [65] includes ATP-dependent RNA helicase DDX1 (CHEMBL2011807), ATP-dependent RNA helicase DDX3X (CHEMBL2011808), Cyclin-T1 CCNT1 (CHEMBL2348842), and Tyrosine-protein kinase HCK (CHEMBL2408778)[65], among other targets. In addition, HVB-related targets are Nuclear receptor subfamily 5 group A member 2 NR5A2 (CHEMBL3544), Interferon-induced, double-stranded RNA-activated protein kinase EIF2AK2 (CHEMBL5785) and SRSF protein kinase 1 SRPK1 (CHEMBL4375). We hope that this list can be the starting point for further pharmaceutical studies.

**Phase-separating proteins are enriched in the 5' interactions**

A number of proteins identified in our *cat*RAPID calculations have been previously reported to coalesce in large ribonucleoprotein assemblies such as stress granules. Among these proteins, we found double-stranded RNA-activated protein kinase EIF2AK2 (P19525), Nucleolin NCL (P19338), ATP-dependent RNA helicase DDX1 (Q92499), Cyclin-T1 CCNT1 (O60563), signal recognition particle subunit SRP72 (O76094), LARP7 (Q4G0J3) and La-related protein 4B LARP4B (Q92615) as well as Polypyrimidine tract-binding protein 1 PTBP1 (P26599) and Heterogeneous nuclear ribonucleoprotein Q HNRNPQ (O60506) [63]. To further investigate the propensity of these proteins to phase separate in stress granules, we used the *cat*GRANULE algorithm (**Materials and Methods**) [66] . We found that the 274 proteins binding to the 5' (fragment 1) with Z score ≥1.5 are highly prone to accumulate in stress-granules (274 proteins with the lowest Z score are used in the comparison; p-value<0.0001; Kolmogorov-Smirnoff; **Fig. 4G; Supp. Table 4**).

This finding is particularly relevant because RNA viruses are known to antagonize stress granules formation [53]. Indeed, the role of stress granules and processing bodies in translation suppression and RNA decay have impact on virus replication [67].

## CONCLUSIONS

Our study is motivated by the need to identify molecular mechanisms involved in Covid-19 spreading. Using advanced computational approaches, we investigated the structural content of SARS-CoV-2 RNA and predicted human proteins that bind to it.

We employed *CROSS* [13,68] to compare the structural properties of 2800 coronaviruses and identified elements conserved in SARS-CoV-2 strains. The regions containing the highest amount of structure are the 5' as well as glycoproteins spike S and membrane M.

We found that the spike S protein domain encompassing amino acids 330-500 is highly conserved across SARS-CoV-2 strains. This result suggests that spike S must have evolved to specifically interact with its host partner ACE2 [30] and mutations increasing the binding affinity are highly infrequent. As the nucleic acids encoding for this region are enriched in double-stranded content, we speculate that the structure might attract host regulatory elements, thus further constraining the variability. The fact that the Spike S region is highly conserved among all the analysed SARS-CoV-2 strains suggests that a specific drug can be designed against it to prevent interactions within the host.

By contrast, the highly variable region at amino acids 243-302 in spike S protein corresponds to the binding site of sialic acids in MERS-CoV (see manuscript by E. Milanetti *et al.*) [7,10,34] and could play a role in infection [33]. The fact that the binding region changes in the different strains might indicate a variety of binding affinities for sialic acids, which could provide clues on the specific responses in the human population. Interestingly, the sialic acid binding site is absent in SARS-CoV but present in MERS-CoV, which represents an important difference between the diseases.

Both our sequence and structural analyses of spike S protein indicate high conservation among coronaviruses and suggest that human engineering of SARS-CoV-2 is highly unlikely.

Using *cat*RAPID [18,19] we computed >100000 protein interactions with SARS-CoV-2 and found previously reported interactions such as Polypyrimidine tract-binding protein 1 PTBP1, Heterogeneous nuclear ribonucleoprotein Q HNRNPQ and Nucleolin NCL [43]. In addition, we discovered that the highly structured region at the 5' has the largest number of protein partners including ATP-dependent RNA helicase DDX1 that was previously reported to be essential for HIV-1 and coronavirus IBV replication [54,55], and the double-stranded RNA-specific editases ADAR and ADARB1 that catalyse the hydrolytic deamination of adenosine to inosine [49] . Other predicted interactions are XRCC5 and XRCC6 members of the HDP-RNP complex associating with ATP-dependent RNA helicase DHX9 [69] as well as and 2-5A-dependent ribonuclease RNASEL and 2-5-oligoadenylate synthase 2 OAS2 that control viral RNA degradation [70,71]. Interestingly, DDX1, XRCC6 and OAS2 are upregulated in human alveolar adenocarcinoma cells infected with SARS-CoV-2 [40]. In agreement with our predictions, recent experimental work indicates that the family of ADAR deaminases is active in bronchoalveolar lavage fluids derived from SARS-CoV-2 patients [42].

A significant overlap exists with the list of protein interactions reported by Gordon *et al.* [58], and among the candidate partners we identified AKAP8L, involved as a DEAD/H-box RNA helicase binding protein involved in HIV infection [59]. In general, proteins associated with retroviral replication are expected to play different roles in SARS-CoV-2. As SARS-CoV-2 massively represses host gene expression [49], we hypothesize that the virus hijacks host pathways by recruiting transcriptional and post-transcriptional elements interacting with polymerase II genes and splicing factors such as for instance A-kinase anchor protein 8-like AKAP8L and La-related protein 7 LARP7 that is upregulated in human alveolar adenocarcinoma cells infected with SARS-CoV-2 [40]. The link to proteins previously studied in the context of HIV and other viruses, if further confirmed fro, is particularly relevant for the repurposing of existing drugs [65].

The idea that SARS-CoV-2 sequesters different elements of the transcriptional machinery is particularly intriguing and is supported by the fact that a large number of proteins identified in our screening are found in stress granules [63]. Indeed, stress granules protect the host innate immunity and are hijacked by viruses to favour their own replication [67]. Moreover, as coronaviruses transcription uses discontinuous RNA synthesis that involves high-frequency recombination [43], it is possible that pieces of the viruses resulting from a mechanism called defective interfering RNAs [72] could act as scaffold to attract host proteins [14,15].

**Acknowledgements**

The authors would like to thank Dr. Mattia Miotto, Dr Lorenzo Di Rienzo, Dr. Alexandros Armaos, Dr. Alessandro Dasti and Dr. Claudia Giambartolomei for discussions. We are particularly grateful to Prof. Annalisa Pastore for critical reading, Dr. Gilles Mirambeau for the RT vs RdRP analysis, Dr. Andrea Cerase for the discussing on stress granules and Dr. Roberto Giambruno for pointing to PTBP1 and HNRNPQ experiments.

The research leading to these results has been supported by European Research Council (RIBOMYLOME_309545 and ASTRA_855923), the H2020 projects IASIS_727658 and INFORE_825080, the Spanish Ministry of Economy and Competitiveness BFU2017-86970-P as well as the collaboration with Peter St. George-Hyslop financed by the Wellcome Trust.

**Contributions.** GGT and RDP conceived the study. AV carried out *cat*RAPID analysis of protein interactions, RDP calculated *CROSS* structures of coronaviruses, GGT, MM and EM performed and analysed sequence alignments, JR, EZ and EB analysed the prediction results. AV, RDP and GGT wrote the paper.

**The authors do not have conflicts of interests.**

## MATERIALS AND METHODS

### *Structure prediction*

We predicted the secondary structure of transcripts using *CROSS* (Computational Recognition of Secondary Structure [13,68]. *CROSS* was developed to perform high-throughput RNA profiling. The algorithm predicts the structural profile (single- and double-stranded state) at single-nucleotide resolution using sequence information only and without sequence length restrictions (scores > 0 indicate double stranded regions). We used the *Vienna* method [27] to further investigate the RNA secondary structure of minima and maxima identified with *CROSS* [13].

### *Structural conservation*

We used *CROSS*align [13,68] an algorithm based on Dynamic Time Warping (DTW), to check and evaluate the structural conservation between different viral genomes [13]. CROSS*align* was previously employed to study the structural conservation of ~5000 HIV genomes. SARS-CoV-2 fragments (1000 nt, not overlapping) were searched inside other complete genomes using the OBE (open begin and end) module, in order to search a small profile inside a larger one. The lower the structural distance, the higher the structural similarities (with a minimum of 0 for almost identical secondary structure profiles). The significance is assessed as in the original publication [13].

### *Sequence collection*

The FASTA sequences of the complete genomes of SARS-CoV-2 were downloaded from Virus Pathogen Resource (VIPR; www.viprbrc.org), for a total of 62 strains. Regarding the overall coronaviruses, the sequences were downloaded from NCBI selecting only complete genomes, for a total of 2862 genomes. The reference Wuhan sequence with available annotation (EPI_ISL_402119) was downloaded from Global Initiative on Sharing All Influenza Data. (GISAID https://www.gisaid.org/).

***Protein-RNA interaction prediction***

Interactions between each fragment of target sequence and the human proteome were predicted using *cat*RAPID *omics* [18,19] , an algorithm that estimates the binding propensity of protein-RNA pairs by combining secondary structure, hydrogen bonding and van der Waals contributions. As reported in a recent analysis of about half a million of experimentally validated interactions [31], the algorithm is able to separate interacting vs non-interacting pairs with an area under the ROC curve of 0.78. The complete list of interactions between the 30 fragments and the human proteome is available at http://crg-webservice.s3.amazonaws.com/submissions/2020-03/252523/output/index.html?unlock=f6ca306af0. The output then is filtered according to the Z-score column, which is the interaction propensity normalised by the mean and standard deviation calculated over the reference RBP set (http://s.tartaglialab.com/static_files/shared/faqs.html#4). We used three different thresholds in ascending order of stringency: Z greater or equal than 1.50, 1.75 and 2 respectively and for each threshold we then selected the proteins that were unique for each fragment for each threshold. *omiXscore* calculations of ADAR and ADARB1 are interactions are respectively at http://crg-webservice.s3.amazonaws.com/submissions/2020-04/263420/output/index.html?unlock=f9375fdbf9 and http://crg-webservice.s3.amazonaws.com/submissions/2020-04/263140/output/index.html?unlock=bb28d715ea.

***GO terms analysis***

*clever*GO [38], an algorithm for the analysis of Gene Ontology annotations, was used to determine which fragments present enrichment in GO terms related to viral processes. Analysis of functional annotations was performed in parallel with *GeneMania* [46].

***RNA and protein alignments***

We used *Clustal W* [28] for 62 SARS-CoV-2 strains alignments and *T-Coffee* [32] for spike S proteins alignments. The variability in the spike S region was measured by computing Shannon entropy on translated RNA sequences. The Shannon entropy is computed as follows:

S(a) = - Sum_i P(a,i) log P(a,i)

Where *a* correspond to the amino acid at the position *i* and P(a,i) is the frequency of a certain amino-acid *a* at position *i* of the sequence. Low entropy indicates poorly variability: if P(a,x) = 1 for one *a* and 0 for the rest, then S(x) =0. By contrast, if the frequencies of all amino acids are equally distributed, the entropy reaches its maximum possible value.

***Predictions of phase separation***

*cat*GRANULE [66] was employed to identify proteins assembling into biological condensates. Scores > 0 indicate that a protein is prone to phase separate. Structural disorder, nucleic acid binding propensity and amino acid patterns such as arginine-glycine and phenylalanine-glycine are key features combined in this computational approach [66].

**FIGURES LEGENDS**

***Fig. 1.*** *Using the CROSS approach [13,68], we studied the structural content of SARS-CoV-2. We found the highest density of double-stranded regions in the 5' (nucleotides 1-253), membrane M protein (nucleotides 26523-27191), and the spike S protein (nucleotides 23000-24000). Strong match is observed between CROSS and Vienna analyses (centroid structures shown), indicating that regions with the highest structural content have the lowest free energies.*

***Fig. 2.*** *We employed the CROSSalign approach [13,68] to compare the Wuhan strain MN908947 with other coronaviruses (1387 strains, including SARS-CoV and MERS-CoV) indicates that the most conserved region falls inside the spike S genomic locus. The inset shows thermodynamic structural variability (positional entropy) within regions encompassing nucleotides 23000-24000 along with the centroid structure and free energy.*

***Fig. 3.*** *Sequence and structural comparison of human SARS-CoV-2 strains. (**A**) Strong sequence conservation (Clustal W multiple sequence alignments [38]) is observed in coding regions, including the region between nucleotides 23000 and 24000 of spike S protein. High structural variability (red bars on top) is observed for both the UTRs and for nucleotides between 21000 and 22000 as well as 24000 and 25000, associated with the S region. The rest of the regions are significantly conserved at a structural level. (**B**) The sequence variability (Shannon entropy computed on T-Coffee multiple sequence alignments [32]) in the spike S protein indicate conservation between amino-acids 460 and 520 (blue box) binding to the host receptor angiotensin-converting enzyme 2 ACE2. The region encompassing amino-acids 243 and 302 is highly variable and is implicated in sialic acids in MERS-CoV (red box). The S1 and S2 domains of Spike S protein are displayed.*

***Fig. 4.*** *Characterization of protein interactions with SARS-CoV-2 RNA. (**A**) In agreement with studies on coronaviruses [35], PTBP1 shows the highest interaction propensity at the 5' and HNRNPQ at 3' (indicated by red bars). (**B**) Number of RBP interactions for different SARS-CoV-2 regions (colours indicate catRAPID [18,19] confidence levels: Z=1.5 or low Z=1.75 or medium and Z=2.0 or high; regions with scores lower than Z=1.5 are omitted); (**C**) Enrichment of viral processes in the 5' of SARS-CoV-2 (precision = term precision calculated from the GO graph structure lvl = depth of the term; go_term = GO term identifier, with link to term description at AmiGO website; description = label for the term; e/d = enrichment / depletion compared to the population; %_set = coverage on the provided set; %_pop = coverage of the same term on the*

*population; p_bonf = p-value of the enrichment. To correct for multiple testing bias, use Bonferroni correction)* [38]*; (**D**) Viral processes are the third largest cluster identified in our analysis; (**E**) Protein interactions with the 5' of SARS-CoV-2 RNA (inner circle) and associations with other human genes retrieved from literature (green: genetic associations; pink: physical associations); (**F**) Number of RBP interactions identified by Gordon et al.* [58] *for different SARS-CoV-2 regions (see panel A for reference). (**G**) Proteins binding to the 5' with Z score ≥ 1.5 show high propensity to accumulate in stress-granules (same number of proteins with Z score* < -1.5 *are used in the comparison; *** p-value<0.0001; Kolmogorov-Smirnoff).*

**SUPPLEMENTARY MATERIAL**

***Supp. Figure 1.*** *We employed CROSSalign [13,68] was to compare the Wuhan strain MN908947 with other coronaviruses (2800 strains, including SARS-CoV, MERS-CoV and coronaviruses having as host other species, such as bats). The result highlights that the most conserved region falls inside the spike S genomic locus.*

***Supp. Table 1.*** *1) catRAPID [18,19] score for interactions with fragment 1; 2) GO [38] and Uniprot annotations of viral proteins interacting with fragment 1 and ; 3**)** catRAPID score for interactions with fragment 2; 4) GO annotations of viral proteins interacting with fragment 2; 5**)** catRAPID score for interactions with fragment 29; 6) GO annotations of viral proteins interacting with fragment 29;*

***Supp. Table 2.*** *RBP interactions from Gordon et al. [58] classified according to catRAPID scores. GO [38] and Uniprot [39] annotations are reported.*

***Supp. Table 3***. *RBPs significantly enriched in the 5' with GO [38] and Uniprot [39] annotations for HIV, HBV and Influenza.*

***Supp. Table 4***. *Stress granule propensity computed with catGRANULE [66] for proteins with interactions score $Z \geq 1.5$ (top-ranked) and $Z<-1.5$ (bottom-ranked). The two groups of proteins have the same number of elements.*

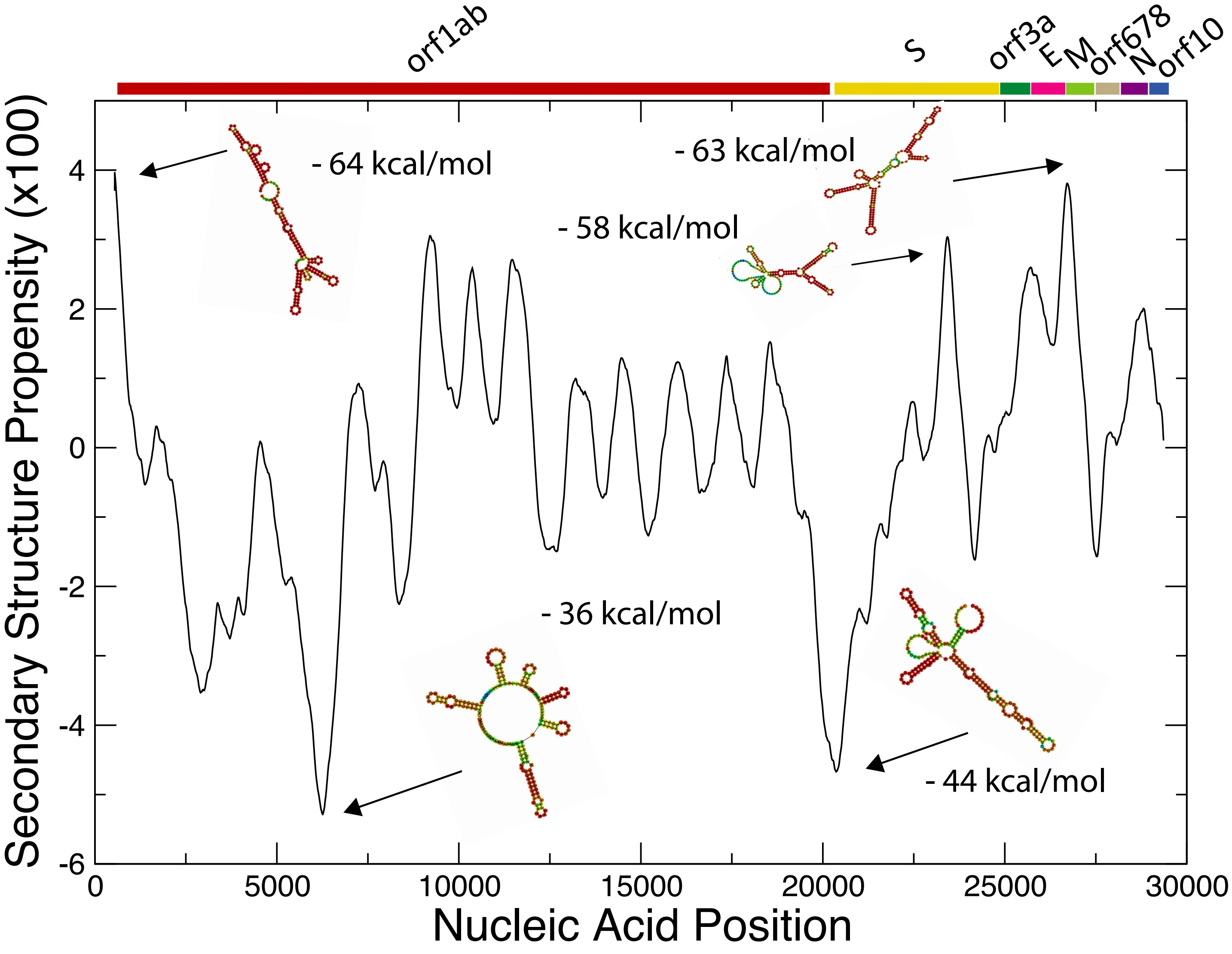

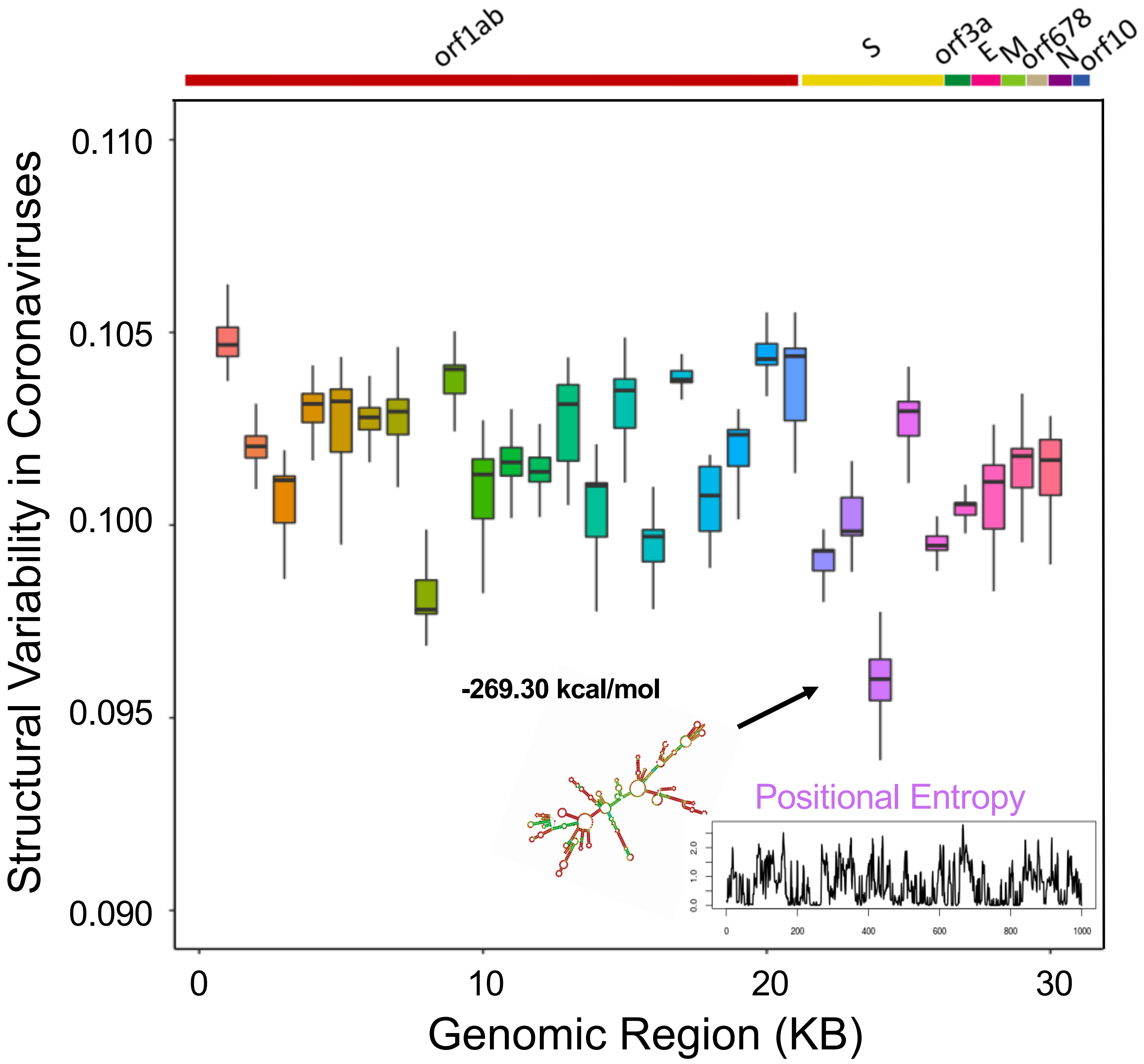

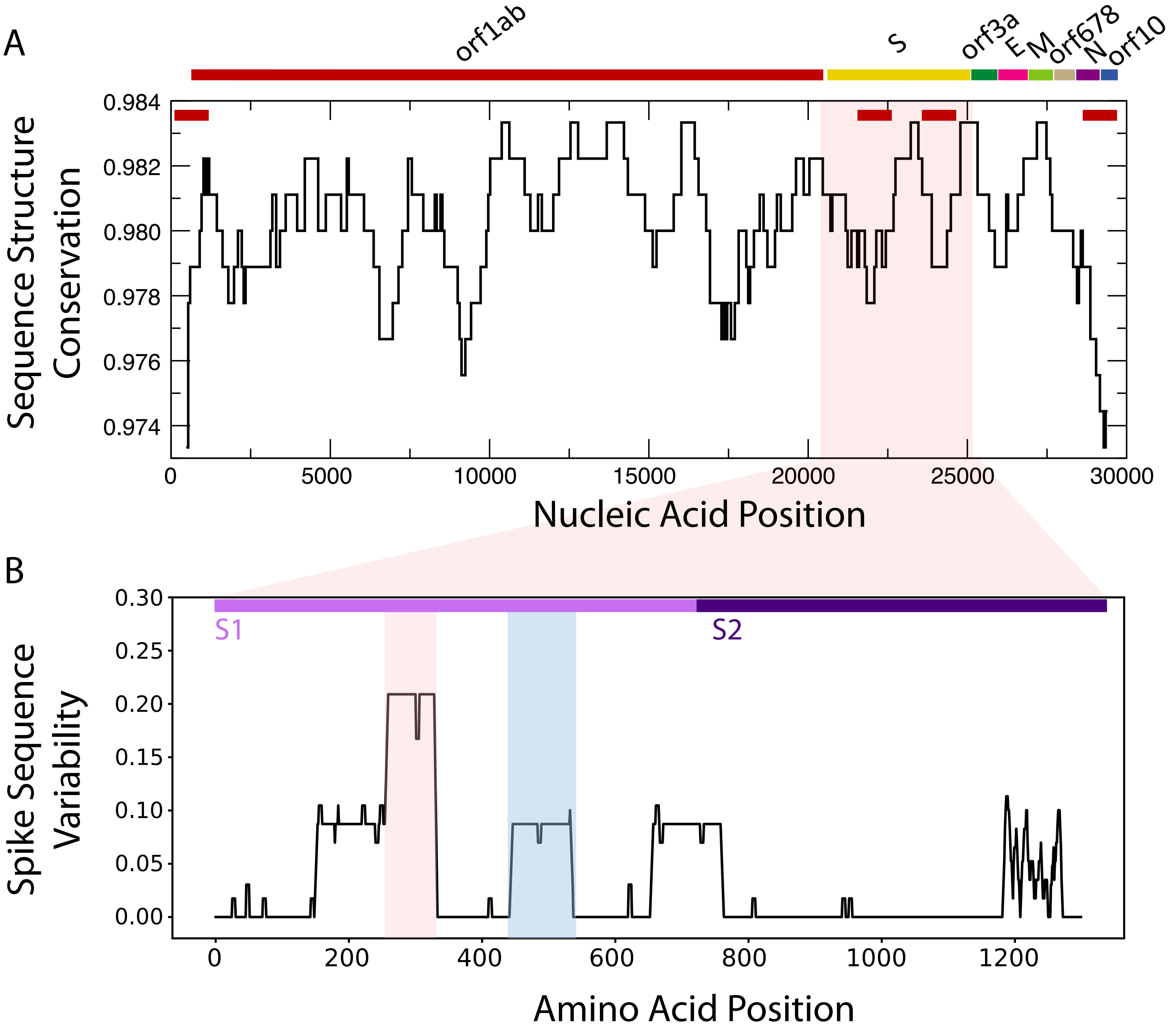

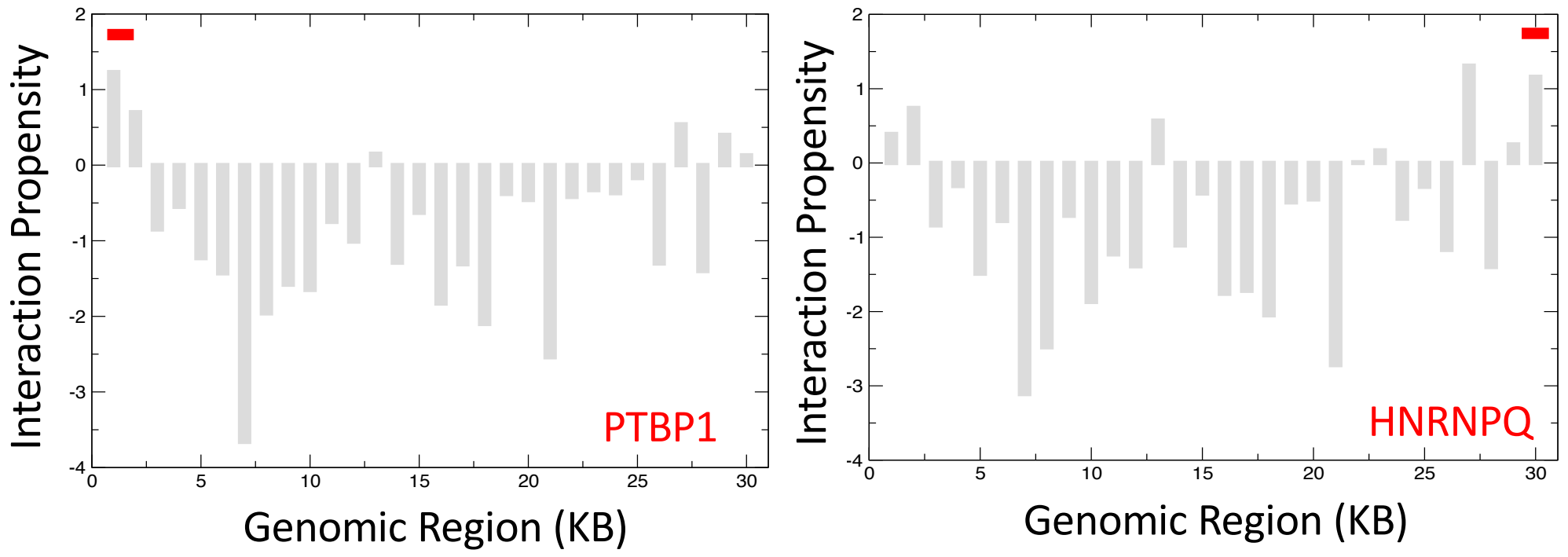

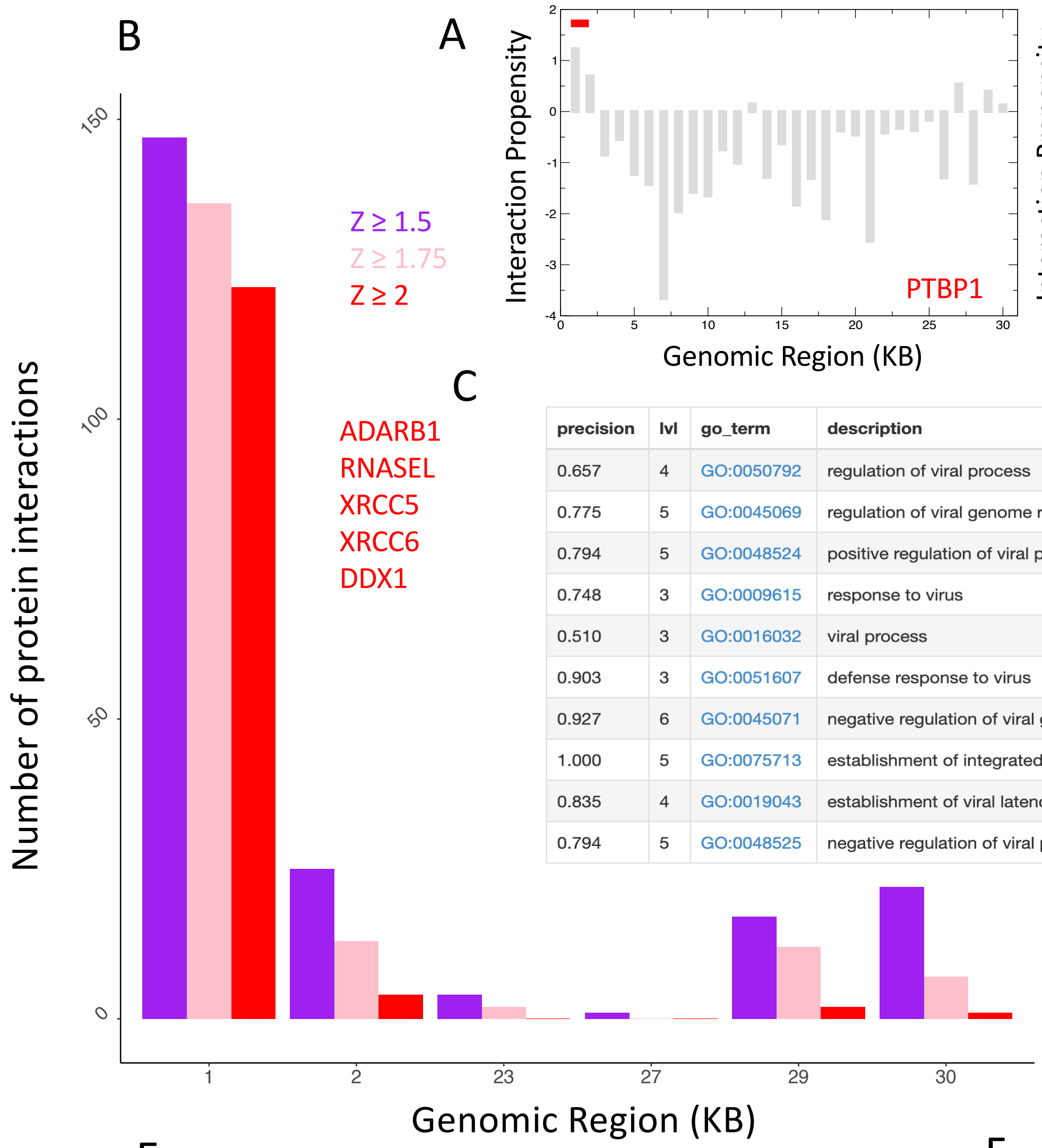

C

| precision | lvl | go_term | description | e/d | %_set | %_pop | p_bonf |
|---|---|---|---|---|---|---|---|
| 0.657 | 4 | GO:0050792 | regulation of viral process | e | 4.918 | 0.205 | 4.59e-4 |
| 0.775 | 5 | GO:0045069 | regulation of viral genome replication | e | 3.279 | 0.083 | 7.33e-3 |
| 0.794 | 5 | GO:0048524 | positive regulation of viral process | e | 3.279 | 0.091 | 1.05e-2 |
| 0.748 | 3 | GO:0009615 | response to virus | e | 4.918 | 0.362 | 1.23e-2 |
| 0.510 | 3 | GO:0016032 | viral process | e | 6.557 | 0.791 | 1.3e-2 |
| 0.903 | 3 | GO:0051607 | defense response to virus | e | 4.098 | 0.216 | 1.53e-2 |
| 0.927 | 6 | GO:0045071 | negative regulation of viral genome replication | e | 2.459 | 0.053 | 7.93e-2 |
| 1.000 | 5 | GO:0075713 | establishment of integrated proviral latency | e | 1.639 | 0.009 | 1.05e-1 |
| 0.835 | 4 | GO:0019043 | establishment of viral latency | e | 1.639 | 0.011 | 1.69e-1 |
| 0.794 | 5 | GO:0048525 | negative regulation of viral process | e | 2.459 | 0.098 | 4.98e-1 |

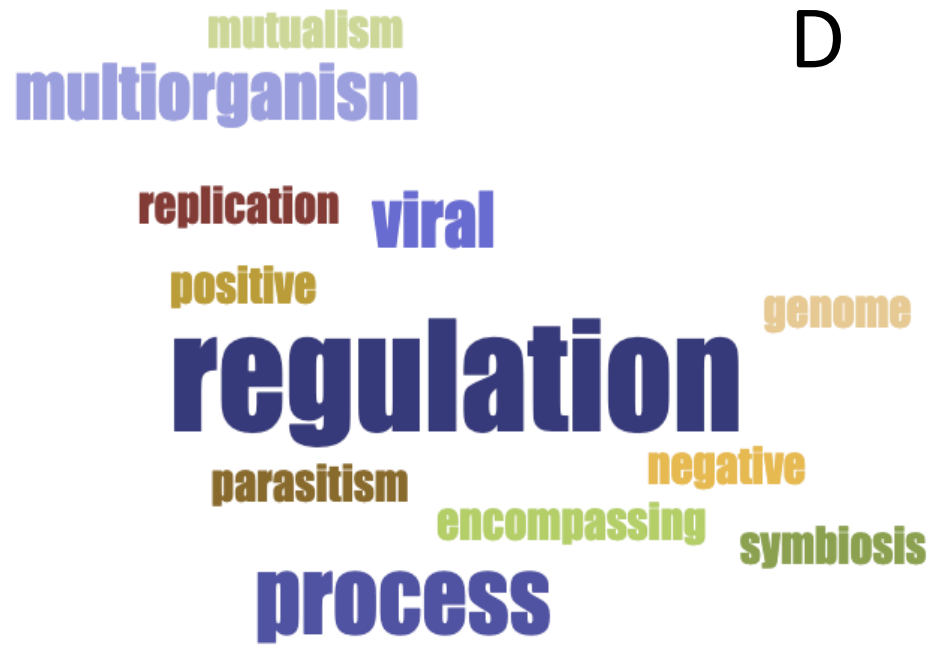

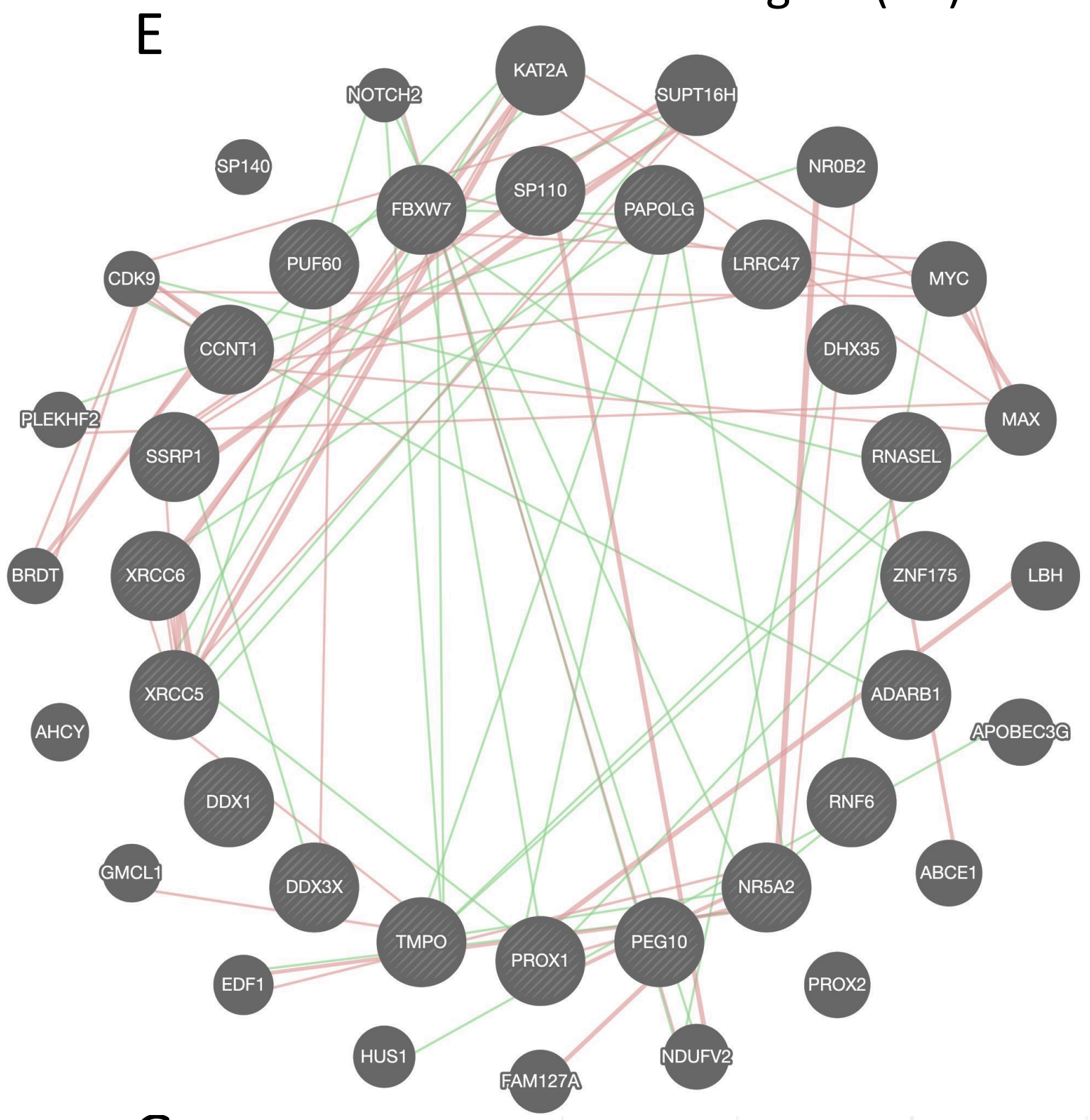

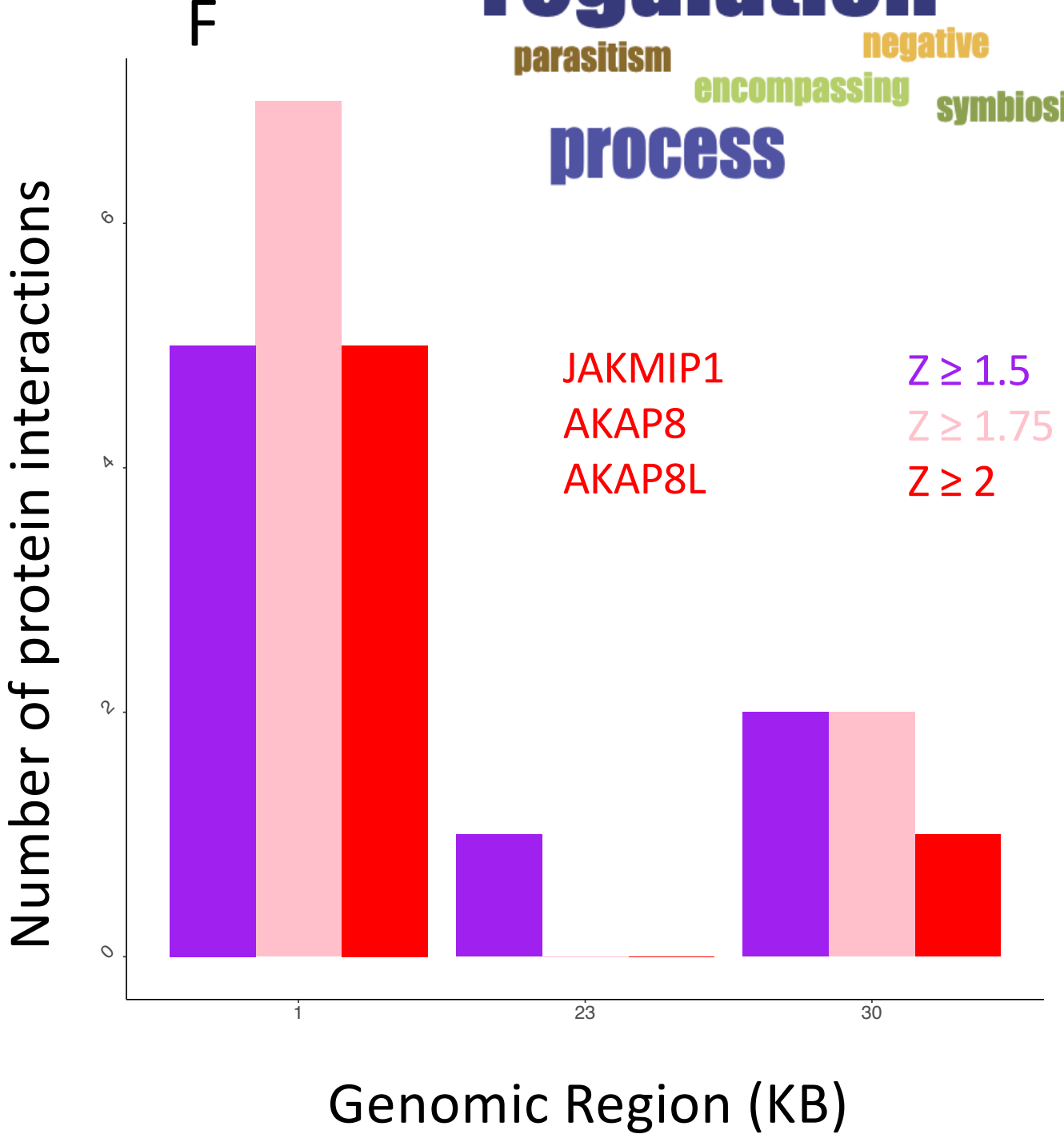

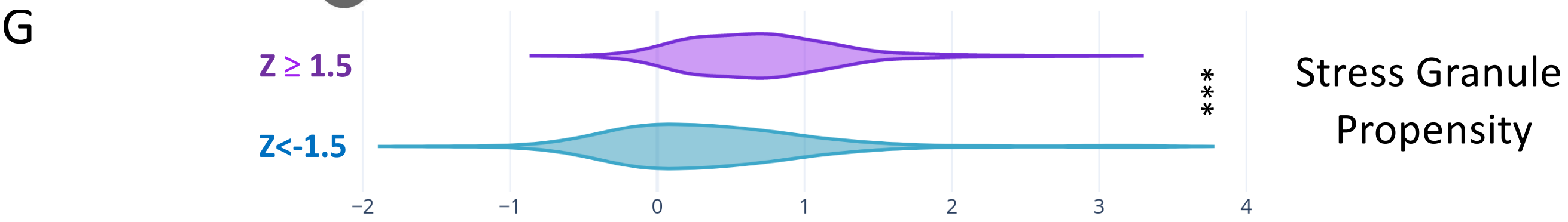